\documentclass{elsart}
\usepackage{natbib}
\usepackage{graphics}
\usepackage{amsmath}
\begin{document}
  \begin{frontmatter}
    \title{The particle-in-cell model for {\em ab initio} thermodynamics:
      implications for the elastic anisotropy of the Earth's
      inner core.}
    
    \author[UCL]{C. M. S. Gannarelli}
    \author[UCL,Bbk]{D. Alf\`{e}}
    \author[UCL]{M. J. Gillan\corauthref{mjg}}
    \corauth[mjg]{Corresponding Author.
      Tel: +44 (0)20 7679 7049;
      Fax: +44 (0)20 7679 1360}
    \ead{m.gillan@ucl.ac.uk}
    \ead[url]{http://www.cmmp.ucl.ac.uk/~mjg}
    \address[UCL]{%
      Physics and Astronomy Department,
      University College London,
      Gower Street,
      London WC1E 6BT,
      United Kingdom}
    \address[Bbk]{%
      Research School of Geological and Geophysical Sciences,
      Birkbeck and University College London,
      Gower Street,
      London WC1E 6BT,
      United Kingdom}
    \begin{abstract}
      We assess the quantitative accuracy
      of the particle-in-cell (PIC) approximation used in
      recent {\em ab initio\/} predictions of the thermodynamic
      properties of hexagonal-close-packed iron at the
      conditions of the Earth's inner core. The assessment
      is made by comparing PIC predictions for a range
      of thermodynamic properties with the results of more
      exact calculations that avoid the PIC approximation. It
      is shown that PIC gives very accurate results for some
      properties, but that it gives an unreliable treatment of
      anharmonic lattice vibrations. In addition, our assessment
      does not support recent PIC-based predictions that the
      hexagonal \(c/a\) ratio increases strongly with increasing
      temperature, and we point out that this casts doubt on
      a proposed re-interpretation of the elastic anisotropy 
      of the inner core. 
    \end{abstract}
    \begin{keyword}
      {\em Ab initio} \sep Anisotropy \sep Core \sep Thermodynamics
      \PACS 62.20.D \sep 62.50.+p \sep 71.15.Ap \sep 91.35.-x
    \end{keyword}
  \end{frontmatter}
  
  \section{Introduction}
  {\em Ab initio\/} calculations based on density functional
  theory (DFT) (\citet{hk:1964,ks:1965,jg:1989}) are widely
  used to calculate the properties of materials
  at the extreme pressures found in the interior
  of the Earth and other planets (\citet{sch:1998}),
  and are known to be capable of high accuracy. For
  many years, such calculations did not generally
  include thermal effects, and explicitly treated
  only zero-temperature materials. However, the
  past few years have seen an increasing effort
  to apply DFT calculations to high-temperature
  solids and liquids of geological interest, using
  {\em ab initio\/} lattice dynamics or molecular-dynamics
  simulation (\citet{apg:2001,apg:2002,baj:2000,lbcst:2000,%
    swc:1997,wsc:1996,bop:2002,obp:2001,ssc:2002}).
  Because dynamical DFT calculations
  demand large computer resources, a simplifying
  approximation known as the `particle-in-cell' (PIC) method
  (\citet{hcb:1954,hr:1970,h_ea:1970,rh:1973,wc:1975,c_ea:1990})
  has been used in some work on
  high-\(T\), high-\(p\) solids, an important 
  example being the very recent use of the PIC
  approximation to re-interpret the elastic anisotropy
  of the Earth's solid inner core (\citet{sscg:2001,ssc:2002}).  However, the
  reliability of the PIC method has not gone
  unchallenged (\citet{apg:2001}) and conclusions based on it are not
  necessarily secure. To shed light on this
  question, we present here the results of our
  own PIC calculations of a range of thermodynamic
  properties of solid iron at Earth's core conditions,
  which we compare with more exact calculations
  that avoid the PIC approximation.
  
  The purpose of the PIC approximation is to provide a
  way of calculating the free energy of a vibrating 
  crystal. The essence of the approximation is that
  correlations between the vibrational displacements
  of different atoms are neglected, so that each
  atom is treated as vibrating independently of every
  other, as in the Einstein model for a vibrating
  solid. Although the approximation seems at first
  sight rather crude, empirically it has been shown
  to yield satisfactory predictions for the
  thermodynamic properties of a number of solids
  (\citet{hr:1970,h_ea:1970,rh:1973,wc:1975}),
  at least for temperatures above the Debye temperature.
  In the geological context, it was shown by \citet{wsc:1996} that,
  when implemented using DFT methods, it gives
  predictions for the \(p(V)\) relation on the shock
  Hugoniot in excellent agreement with experiment for
  Fe up to the melting curve. The PIC approach based
  on DFT has also been
  applied successfully to the high-\(p\), high-\(T\)
  properties of other transition metals
  (\citet{cg_tant:2001,gc_tant:2002}).
  
  In the work on the elastic anisotropy of
  the Earth's inner core (\citet{sscg:2001,ssc:2002}) mentioned above, 
  PIC was used to calculate the elastic constants
  of hexagonal-close-packed (hcp) Fe over a range
  of pressures and temperatures. A key prediction was
  that the \(c/a\) ratio increases strongly with
  \(T\), and that this leads in turn to a strong
  \(T\)-dependence of the elastic constants. It emerged
  from this that the observed anisotropy of seismic
  velocities in the inner core (\citet{creager:1992,sh:1993,tromp:1993})
  could be interpreted
  in terms of a partial alignment of hcp crystallites.
  Remarkably, the calculations required the hcp basal
  plane to be preferentially aligned along the Earth's
  rotational axis, which is the opposite
  of what had been proposed earlier (\citet{sc:1995,ssc:1999}).
  Since the correctness of this interpretation depends 
  heavily on the predicted variation of
  \(c/a\) with \(T\), it is clearly essential to be
  confident that the PIC approximation can
  be relied on to give this variation correctly,
  and one of our aims in this paper is to test
  this point.
  
  An objective assessment of the errors incurred by
  the PIC approximation is made possible by
  the fact that {\em ab initio\/} free energies
  and other thermodynamic functions of
  high-temperature solids can now be calculated with
  statistical-mechanical errors that can be made
  arbitrarily small. The new methods, described in
  detail in previous papers (\citet{apg:2001,apg:2002}),
  are based on the DFT
  calculation of phonon frequencies in the harmonic
  approximation, supplemented by the `thermodynamic
  integration' technique (\citet{frenk:1996}) for calculating anharmonic
  contributions. For a given material, and with a
  given DFT technique for calculating the
  electronic total energy as a function of
  ionic positions, we can therefore compute
  thermodynamic functions either with the
  PIC approximation or almost exactly. The 
  differences between the two sets of results
  represent the errors caused by PIC. This is
  what we do in the present work.
  Since the main difference between PIC and the
  newer, more exact methods is that the latter fully
  include vibrational correlations, we shall
  refer to these in the following as `vibrationally
  correlated' calculations.
  
  Two main claims have been made for the PIC
  method (\citet{wsc:1996}): First, that it provides a simple
  and reasonably accurate way of including the
  effect of the anharmonicity of lattice vibrations
  on thermodynamic properties; and 
  second, that, even in the absence of 
  anharmonicity, it is computationally
  much less demanding than more precise methods.
  Our assessment of these claims for the case of
  high-\(p\)/high-\(T\) Fe will suggest that neither is
  necessarily true, but that nevertheless the PIC method does
  yield surprisingly accurate predictions for
  many thermodynamic properties. We shall
  elucidate the reasons why PIC is often accurate.
  As we shall see, however, the 
  present work does not support
  the prediction that the \(c/a\) ratio of hcp Fe
  increases strongly with temperature, and this
  casts doubt on the proposed re-interpretation
  of the elastic anisotropy of the Earth's inner
  core.
  
  The remainder of the paper is organised as follows.
  In Sec. 2, we describe how we have applied PIC
  to hcp Fe and how we have separated the various
  contributions to the free energy and other
  thermodynamic functions; at the end of Sec. 2,
  we summarise the DFT techniques. In
  Sec. 3, we describe the details of our
  free-energy calculations and highlight their
  implications for the temperature dependence of
  the \(c/a\) ratio. We then present results for
  several thermodynamic quantities calculated in
  the PIC approximation, which we compare with the
  earlier PIC results of Cohen and co-workers
  (\citet{swc:1997,sscg:2001,ssc:2002,wsc:1996}) and with the
  vibrationally correlated results of \citet{apg:2001}.
  Discussion and conclusions follow in Sec. 4.
  
  \section{Techniques}
  \subsection{Particle-in-cell approximation}
  \label{pic}
  In classical statistical mechanics, the {\em ab initio\/} Helmholtz
  free energy of a vibrating solid containing \(N\) ions is:
  \begin{equation}
    \label{full_f}
    F_{\rm AI} = - k_{\rm B} T \ln \left\{ \frac{1}{\Lambda^{3 N}}
    \int d {\bf r}_1 \ldots d {\bf r}_N \exp \left[ - \beta U_{\rm AI}
      ( {\bf r}_1 \, \ldots {\bf r}_N ) \right] \right\} \; ,
  \end{equation}
  where \(U_{\rm AI} ( {\bf r}_1 \, \ldots {\bf r}_N )\) is the {\em ab initio\/}
  total electronic free energy of the system when the ionic positions
  are \({\bf r}_1 , \ldots {\bf r}_N\) and \(\Lambda\) is the thermal 
  wavelength. It should be noted that \(F_{\rm AI}\) depends on the
  volume \(V\) and temperature \(T\) of the system (\(\beta =
  1 / k_{\rm B} T\)), and for an hcp crystal it also depends on the
  axial \(c/a\) ratio, denoted here by \(q\); for the moment we do not indicate
  explicitly the dependence on \(V\), \(T\) and \(q\).
  Two other points should also be recalled.  First, the quantity
  \(U_{\rm AI}({\bf r}_1,\ldots,{\bf r}_N)\) is a {\em free} energy,
  because the electrons are treated as being in thermal equilibrium
  for each set of ionic positions, at a temperature equal to the
  temperature \(T\) of the system as a whole. Second, the use of
  {\em classical\/} statistical mechanics, i.e.\ the neglect of quantum
  effects in the nuclear motions, is fully justified at Earth's core
  temperatures (\citet{apg:2001}).
  
  The PIC approach consists in replacing \(U_{\rm AI}\) in the above
  formula by the approximate form \(U_{\rm AI}^{\rm PIC}\), given by:
  \begin{equation}
    \label{def_pic}
    U_{\rm AI}^{\rm PIC} ( {\bf r}_1, \ldots {\bf r}_N ) =
    U_{\rm perf} + \sum_{i = 1}^N \phi ( {\bf u}_i ) \; .
  \end{equation}
  Here, \(U_{\rm perf} \equiv U_{\rm AI} ( {\bf R}_1 , \ldots {\bf R}_N )\)
  is the {\em ab initio\/} free energy of the system when all ions are at their
  perfect-lattice positions \({\bf R}_i\), and \(\phi ( {\bf u}_i )\) is 
  defined to be the change of energy 
  \(U_{\rm AI} ( {\bf R}_1 , \ldots {\bf R}_i + {\bf u}_i , \ldots {\bf R}_N ) -
  U_{\rm perf}\) when ion \(i\) is displaced from its perfect-lattice position
  to the position \({\bf R}_i + {\bf u}_i\), all other ions being held
  fixed at their perfect-lattice positions.
  With \(U_{\rm AI}\) replaced by \(U_{\rm AI}^{\rm PIC}\) in
  Eqn. (\ref{full_f}), the PIC approximation for the free energy is:
  \begin{equation}
    F_{\rm AI}^{\rm PIC} = U_{\rm perf} + N f_{\rm vib}^{\rm PIC} \; ,
  \end{equation}
  where the PIC vibrational free energy per atom is:
  \begin{equation}
    \label{f_vib}
    f_{\rm vib}^{\rm PIC} = - k_{\rm B} T \ln \left[
      \Lambda^{-3} \int d {\bf u} \, \exp ( - \beta \phi ( {\bf u} ) ) \right] \; .
  \end{equation}
  The problem is thus reduced to the calculation of a three-dimensional
  integral, which can be performed numerically.
  
  Our later analysis depends on a clear separation of harmonic
  and anharmonic contributions to \(f_{\rm vib}^{\rm PIC}\), which
  can be accomplished by considering the series expansion of 
  \(\phi ( {\bf u} )\) in powers of ionic displacement \({\bf u}\):
  \begin{equation}
    \phi ( {\bf u} ) = \frac{1}{2!} \sum_{\alpha \beta} \Phi_{\alpha \beta}^{(2)} 
    u_\alpha u_\beta + 
    \frac{1}{3!} \sum_{\alpha \beta \gamma}\Phi_{\alpha \beta \gamma}^{(3)} 
    u_\alpha u_\beta u_\gamma 
    +\frac{1}{4!} \sum_{\alpha \beta \gamma \delta}
    \Phi_{\alpha \beta \gamma \delta}^{(4)} 
    u_\alpha u_\beta u_\gamma u_\delta + \ldots \; ,
    \label{phi_series}
  \end{equation}
  where the Greek indices \(\alpha\), \(\beta\)... indicate Cartesian components.
  The value of \(f_{\rm vib}^{\rm PIC}\) when we retain only
  the quadratic part of $\phi ( {\bf u} )$ is the harmonic vibrational
  free energy in the PIC approximation, denoted by \(f_{\rm harm}^{\rm PIC}\).
  This can be expressed as:
  \begin{equation}
    \label{f_harm}
    f_{\rm harm}^{\rm PIC} = k_{\rm B} T \sum_{\nu = 1}^3 \ln ( \hbar \omega_\nu /
    k_{\rm B} T ) = 3 k_{\rm B} T \ln ( \hbar {\bar{\omega}} / k_{\rm B} T ) \; .
  \end{equation}
  Here \(\omega_\nu\) are the three PIC vibrational frequencies,
  given by \(\det | M \omega_\nu^2 \delta_{\alpha \beta} - 
  \Phi_{\alpha \beta}^{(2)} | = 0\),
  with \(M\) the ionic mass. The geometric-mean frequency
  \(\bar{\omega}\) is defined by Eqn. (\ref{f_harm}).
  The anharmonic contribution \(f_{\rm anharm}^{\rm PIC}\) to
  the free energy is the part of \(f_{\rm vib}^{\rm PIC}\) not
  accounted for by \(f_{\rm harm}^{\rm PIC}\), so that
  \(f_{\rm anharm}^{\rm PIC} \equiv
  f_{\rm vib}^{\rm PIC} - f_{\rm harm}^{\rm PIC}\).
  
  The PIC results reported in Sec. 3 were obtained by computing
  \(\phi ( {\bf u} )\) for a set of vector displacements
  \({\bf u}\), and fitting the results using the power series
  expansion of Eqn. (\ref{phi_series}); in practice, for the range of
  displacements
  that occur with appreciable probability at temperatures below the
  melting temperature, we find that an extremely accurate fit is
  obtained if we retain only terms up to quartic order in \({\bf u}\).
  To show what parameters enter this fit, we write this quartic
  polynomial explicitly for the case of hcp symmetry:
  \begin{equation}
    \label{phi_hcp}
    \begin{split}
      \phi ( {\bf u} ) = &
      \frac{1}{2!} \left[ M \omega_a^2 ( u_x^2 + u_y^2 ) +
	M \omega_c^2 u_z^2 \right] 
      + \frac{1}{3!} K^{(3)} ( u_y^3 - 3 u_x^2 u_y ) \\
      & + \frac{1}{4!} \left [ K_a^{(4)} ( u_x^2 + u_y^2 )^2  
	+  K_{\rm mix}^{(4)} u_z^2 ( u_x^2 + u_y^2 )
	+ K_c^{(4)} u_z^4 \right] \; .
    \end{split}
  \end{equation}
  Here, \(u_x\) and \(u_y\) are perpendicular Cartesian components in the
  basal plane, the \(x\)-axis being oriented towards a nearest-neighbour,
  and \(u_z\) is the displacement along the hexagonal axis. The frequencies
  of harmonic vibration in the basal plane and along the hexagonal axis
  are \(\omega_a\) and \(\omega_c\) respectively. This symmetrised polynomial
  expression may be obtained to a given order by writing down all possible
  terms in a polynomial of that order, and retaining only those terms which
  leave the expression invariant under all the point
  symmetry operations of the chosen cell. In the case of hcp symmetry, these
  include  reflection in the hexagonal (\(z=0\)) plane, rotations of
  \(\frac{2\pi}{3}\) about the hexagonal axis, and reflection in the
  \(y\)--\(z\) plane. With the above form of \(\phi\),
  the geometric-mean frequency \({\bar{\omega}}\) given by
  \begin{equation}
    \label{eq_ombar}
    3 \ln  {\bar{\omega}} = 2 \ln \omega_a + \ln \omega_c \; .
  \end{equation}
  
  The vibrational free energy \(f_{\rm vib}^{\rm PIC}\) is obtained
  essentially exactly by numerical evaluation of the integral in
  Eqn. (\ref{f_vib}) on a regular grid.
  In analysing the anharmonic contributions, it is
  useful to note that \(f_{\rm anharm}^{\rm PIC}\) can be expanded as a power
  series in temperature:
  \begin{equation}
    f_{\rm anharm}^{\rm PIC} = d T^2 + O ( T^3 )\; .
    \label{def_d}
  \end{equation}
  For temperatures
  below the melting point, we find that only the term in \(T^2\) is 
  significant, so that in practice the anharmonic contributions are
  completely specified by the coefficient \(d\) as a function of
  volume \(V\) and \(c/a\) ratio \(q\). The following exact
  expression for the coefficient is readily obtained:
  \begin{equation}
    \label{coeff_anh}
    d = K_a^{(4)} / 3 ( M \omega_a^2 )^2 +
    K_{\rm mix}^{(4)} / 12 ( M \omega_a \omega_c )^2 
    + K_c^{(4)} / 8 ( M \omega_c^2 )^2 - \left( K^{(3)} \right)^2 / 3
    ( M \omega_a^2 )^3 \; .
  \end{equation} 
  
  The above methods are used to obtain contributions to the free energy
  at a reference value of \(q\), chosen to be close to the \(T=0\)
  equilibrium value.  We then perform further calculations over a range
  of values of \(q\), to obtain corrections to the free energy components,
  which are parameterised in terms of \(V\) and \(T\).  The details of
  this parameterisation are given in Sec. \ref{res_perf} and \ref{res_vib}.
  Equilibrium values of \(q\) are found by direct minimisation of the total
  free energy.
  
  \subsection{{\em Ab initio\/} methods}
  The DFT electronic-structure techniques used to 
  perform the PIC calculations are essentially identical
  to those described by (\citet{apg:2001,apg:2002}).
  The exchange-correlation functional is the generalised
  gradient approximation (GGA) of Perdew and Wang
  (\citet{wp:1991,perdew:1992}).  The implementation of DFT
  is the projector augmented wave (PAW) scheme
  (\citet{blochl:1994,kj:1999}), with core radii,
  augmentation charge
  radii etc set to the values reported in \citet{akg:2000}.
  As before, all atomic states up to and including
  $3p$-states are treated as core states, but the high-pressure
  response of $3s$- and $3p$-states, known to be important
  at Earth's-core conditions, is included {\em via} an
  effective pair potential; the accuracy of this procedure
  has been demonstrated earlier (\citet{apg:2001}). Thermal
  excitation of electrons, also important at core
  conditions, is treated with the usual finite-temperature
  formulation of DFT (\citet{mermin:1965,gillan:1989,wma:1992}).
  We used a plane-wave cut-off of 300 eV, as in our previous work. All
  calculations were performed with the VASP
  code (\citet{kf_02:1996,kf_01:1996}).
  
  The PIC vibrational potential $\phi ( {\bf u} )$
  (see Eqn. (\ref{def_pic})) should in principle be 
  calculated by displacing a single atom in
  an infinite crystal. Because PAW
  calculations require periodic boundary
  conditions, we must instead use supercell geometry,
  so that the displaced atom has periodic
  images. To adhere to the PIC scheme,
  we must therefore ensure convergence
  of all results with respect to the size of
  the supercell, as described in the following
  Section.
  
  \section{Results}
  \subsection{Free energy of perfect lattice}
  \label{res_perf}
  DFT results for the free energy of the hcp Fe perfect lattice
  $U_{\rm perf} ( V, q, T )$ for the fixed $c/a$ ratio $q$ equal
  to 1.60 were reported earlier (\citet{apg:2001}) for $V$ values from 5.2
  to 11.4 \AA$^3$/atom at temperatures from 200
  to $10^4$ K. At each $T$ value, the results were
  fitted to a Burch-Murnaghan equation of state (\citet{poirier:2000}),
  the parameters of which were then fitted as polynomial
  functions of $T$. We have repeated these calculations
  for the present work, and as expected the results are
  virtually identical to those reported earlier.
  
  To allow for variation of $q$, we have performed
  additional calculations of
  $U_{\rm perf} ( V, q, T )$ for $q$ in
  the range from 1.48 to 1.72, with $V$ going
  from 5.5 to 10.5 \AA$^3$/atom, and
  $T$ going from 2000 to 8000 K. All
  technical parameters, such as the Monkhorst-Pack
  sampling set, were kept the same as in the
  $q = 1.60$ calculations. To represent the
  results, we define the deviation 
  $\Delta U_{\rm perf} ( V, q, T )$ of the
  perfect-lattice free energy from its value
  at a chosen of $q$, denoted by $q_0$:
  $\Delta U_{\rm perf} ( V, q, T ) \equiv
  U_{\rm perf} ( V, q, T ) - U_{\rm perf} ( V, q_0 , T )$.
  In the present case, $q_0$ has the fixed value 1.60 throughout.
  We find that $\Delta U_{\rm perf} ( V, q, T )$ can
  be very accurately represented at all $(V, T )$
  by the quadratic form:
  \begin{equation}
    \Delta U_{\rm perf} ( q ) = \alpha ( q - q_0 ) ( q - q_1 ) \; .
    \label{delta_u}
  \end{equation}
  At each $V$ value, $\alpha$ and $q_1$ can be accurately
  fitted as linear functions of $T$, the coefficients of
  which are in turn fitted linearly in  $V$. These fits give
  a virtually perfect representation
  of the $\Delta U_{\rm perf} ( V, q, T )$ results.
  Writing \(\alpha=a+bT\), \(q_1=r+sT\) and with \(a=a^{(0)}+a^{(1)}V\)
  and similarly for \(b\), \(r\) and \(s\), the numerical values of the
  parameters are 
  \(a^{(0)}=18.39\), \(a^{(1)}=-1.532\),
  \(b^{(0)}=-3.070\times 10^{-4}\), \(b^{(1)}=1.64908\times 10^{5}\),
  \(r^{(0)}=1.63\), \(r^{(1)}=-7.909\times 10^{-3}\),
  \(s^{(0)}=-7.902\times 10^{-6}\) and \(s^{(1)}=1.817\times 10^{-6}\).
  Units of coefficients are such that \(\alpha\) is in eV, \(T\) in K and
  \(V\) in \AA\(^3\).

  The equilibrium value \(q\) of the perfect lattice, denoted by $q_{\rm eq}$,
  is obtained by minimising $\Delta U_{\rm perf} ( V, q, T )$
  with respect to $q$. At $T = 0$, we find a very weak
  dependence of $q_{\rm eq}$ on $V$, going from
  1.590 at 7.0 \AA$^3$/atom to 1.578 at 10.0 \AA$^3$/atom.
  The $T$-dependence of $\Delta U_{\rm perf} (q)$ is
  also very weak, with \(\alpha\) decreasing linearly from
  7.4 to 6.2~eV, and \(q_1\) increasing from 1.59 and 1.62
  as temperature varies from 2000 to 8000~K. If we now use the results
  to predict $q_{\rm eq}$ for the (hypothetical) high-$T$
  perfect lattice, we find a variation of at most 0.04 at a volume of
  10.0 A\(^3\).
  The
  insignificant $V$- and $T$-dependence of $q_{\rm eq}$ is
  noteworthy, because it implies that any significant variation
  of $q_{\rm eq}$ at high temperatures can come only
  from lattice vibrations, to which we now turn.
  
  \subsection{Vibrational free energy}
  \label{res_vib}
  The basal-plane and axial frequencies $\omega_a$
  and $\omega_c$, and the four anharmonic vibrational
  coefficients $K^{(3)}$, $K_a^{(4)}$, $K_{\rm mix}^{(4)}$
  and $K_c^{(4)}$ were calculated as follows.  For
  given values of \(q\) and \(V\), a supercell was
  constructed in which one of the atoms, the `walker'
  is displaced from its equilibrium position.
  The `walker' is given a series of equally spaced
  displacements in the three directions
  \({\bf {\hat u}}_y\), \({\bf {\hat u}}_z\) and
  \(\frac{1}{\surd 2} ({\bf {\hat u}}_x+\bf {{\hat u}}_z)\),
  the maximum displacements \({\bf r}_{\rm max}\)
  in each direction being chosen so that the
  Boltzmann factor 
  \(\exp\left[-\beta \phi ({\bf r}_{\rm max})\right]\approx 0.1\)
  for the maximum temperatures of interest.  Directions
  were selected to allow the terms in \(\phi({\bf r})\)
  (Eqn. (\ref{phi_hcp})) to be determined as simply as
  possible by least squares fitting. Size convergence
  in terms of supercells is essential to ensure that
  the walker cannot `see' its image.
  We discuss the results of a simple convergence test
  below. All the calculations described here are performed at
  a fixed electronic temperature of 6000~K, since
  statistical mechanical considerations will dominate
  the temperature dependence of \(f_{\rm vib}^{\rm PIC}\).
  At this temperature, vibrational coefficients were
  fully converged with respect to  \(k\)-point sampling at each cell size.
  Monkhorst-Pack (\citet{mp:1976}) sampling was used, with
  \(9 \times 9 \times 5\) \(k\)-points for the 8 atom cell,
  \(7 \times 7 \times 5\) \(k\)-points for 16 atoms and
  \(3 \times 3 \times 3\) for 36.
  Calculations were performed over a range of volumes from
  5.5 \AA\(^3\)/atom to 11.5 \AA\(^3\)/atom with a fixed
  axial ratio of \(q_0=1.60\) as before.  Using 5th order
  polynomials to obtain a high quality fit, the quantities
  $K^{(3)}$, $K_a^{(4)}$, $K_{\rm mix}^{(4)}$, $K_c^{(4)}$
  and \(\ln \bar{\omega}\) were then parameterised in terms
  of volume.  Finally, a correction due to relaxation of
  \(q\) is added to the harmonic free energy, as for that
  of the perfect lattice, by performing a parameterisation
  of \(\bar{\omega}\) in terms of \(q\), which we describe
  below.  Parameterisation of higher order \(K\) coefficients
  in terms of \(q\) yielded only negligible corrections to the
  \(p\)--\(V\) curve.
  
\begin{figure}[h]
    \includegraphics{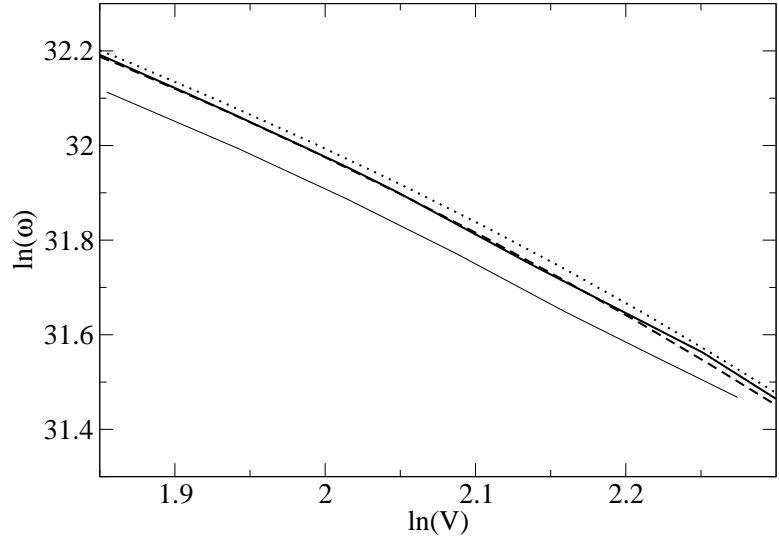}
    \caption{The variation with atomic volume \(V\) of \(\ln \bar{\omega}\)
      (\(\bar{\omega}\) is the geometric mean vibrational frequency) from
      PIC calculation on periodic cells containing 8 (solid curve),
      16 (dashed) and 36 (dotted) atoms.  The vibrationally correlated
      results of \citet{apg:2001} are given by the lighter curve.}
    \label{ombar}
  \end{figure}

  As noted in Sec. \ref{pic}, the harmonic free energy
  is completely determined by the geometric-mean
  frequency $\bar{\omega}$ (Eqns. (\ref{f_harm}) and
  (\ref{eq_ombar})), and we consider first our PIC
  results for $\bar{\omega}$ as a function of $V$
  for the fixed $c/a$ ratio $q_0 = 1.6$. We report
  in Fig. \ref{ombar} our $\bar{\omega}$ results from
  calculations using supercells of 8, 16 and 36 atoms,
  compared with the $\bar{\omega}$ results from
  our earlier calculations of the full phonon
  spectrum (\citet{apg:2001}). We note two important
  points. First, the PIC $\bar{\omega}$ results are
  almost independent of supercell size, so that
  the calculations appear to be fully converged with
  respect to size effects. Second, the PIC $\bar{\omega}$
  differs from the full-phonon $\bar{\omega}$ only by a
  volume-independent shift of $\ln \bar{\omega}$. Since
  such a $V$-independent shift cannot affect most
  thermodynamic quantities, we expect the PIC 
  calculations to agree rather closely with full-phonon
  calculations.
  
  We now consider the dependence of $\bar{\omega}$
  on $c/a$ ratio $q$, defining the logarithmic deviation 
  $\Delta_{\rm L} \bar{\omega} ( V, q )$ from its value
  at $q_0$ by $\ln \bar{\omega} ( V, q ) =
  \ln \bar{\omega} ( V, q_0 ) + \Delta_{\rm L} \bar{\omega} ( V, q )$. 
  In terms of this, the
  harmonic free energy can be written as:
  \begin{equation}
      f_{\rm harm}^{\rm PIC} ( V, q, T ) = 
      3 k_{\rm B} T \ln ( \hbar \bar{\omega} / k_{\rm B} T ) =
      f_{\rm harm}^{\rm PIC} ( V, q_0 , T ) +
      3 k_{\rm B} T \Delta_{\rm L} \bar{\omega} \; .
  \end{equation}
  Calculations of $\bar{\omega} (q)$ were made for
  $q$-values going from 1.48 to 1.72 at a series of
  volumes. We find that $\bar{\omega}$ decreases
  with increasing $q$. Not surprisingly, increase of
  $q$ at constant $V$ yields an increase of the
  basal-plane frequency $\omega_a$ and a decrease
  of axial frequency $\omega_c$. The decrease of
  $\omega_c$ succeeds in outweighing the increase
  of $\omega_a$ in the formula for $\bar{\omega}$
  (Eqn. (\ref{eq_ombar})), but nevertheless the
  resulting decrease of $\bar{\omega}$ is a fairly
  marginal effect. Over the $q$-range studied,
  we found that the quadratic form
  $\Delta_{\rm L} \bar{\omega} (q) = \beta ( q - q_0 ) ( q - q_2 )$
  gives an accurate fit, and this fit was performed at volumes
  of 5.5, 7.0 and 10.5 \AA$^3$/atom. The
  $V$-dependence of the resulting $\beta$
  and $q_2$ coefficients was obtained by a quadratic fit.
  Writing \(\beta=\sum_{i=0}^2\beta^{(i)}V^i\) and similarly
  for \(q_2\), the values of the parameters are
  \(\beta^{(0)}=-2.083\), \(\beta^{(1)}=0.375\), \(\beta^{(2)}=-0.0264\),
  \(q_2^{(0)}=1.990\), \(q_2^{(1)}=-0.141\) and
  \(q_2^{(2)}=8.79\times 10^{-3}\). Units of parameters are such that
  \(\bar{\omega}\) is in rad s\(^{-1}\) and \(V\) is in \AA\(^3\).
    
  From the values of the four anharmonic coefficients
  at each $V$ and $q$, we calculate also the anharmonic
  vibrational free energy $f_{\rm anharm}^{\rm PIC}$.
  As explained in Sec. \ref{pic}, the most accurate way of doing this
  is by explicit calculation of $f_{\rm vib}^{\rm PIC}$
  (Eqn. (\ref{f_vib})) by numerical evaluation of
  the integral over displacement ${\bf  u}$, from which
  $f_{\rm anharm}^{\rm PIC}$ is obtained as the
  difference $f_{\rm vib}^{\rm PIC} - f_{\rm harm}^{\rm PIC}$.
  Alternatively, we can use Eqn. (\ref{coeff_anh}) to calculate the 
  $d$ coefficient of the leading term in the temperature
  expansion of $f_{\rm anharm}^{\rm PIC}$ (see Eqn. (\ref{def_d})).
  In practice, we find that the two methods yield
  almost indistinguishable results, and everything that follows
  is based on evaluation of the $d$ coefficient.
  
\begin{figure}[h]
    \includegraphics{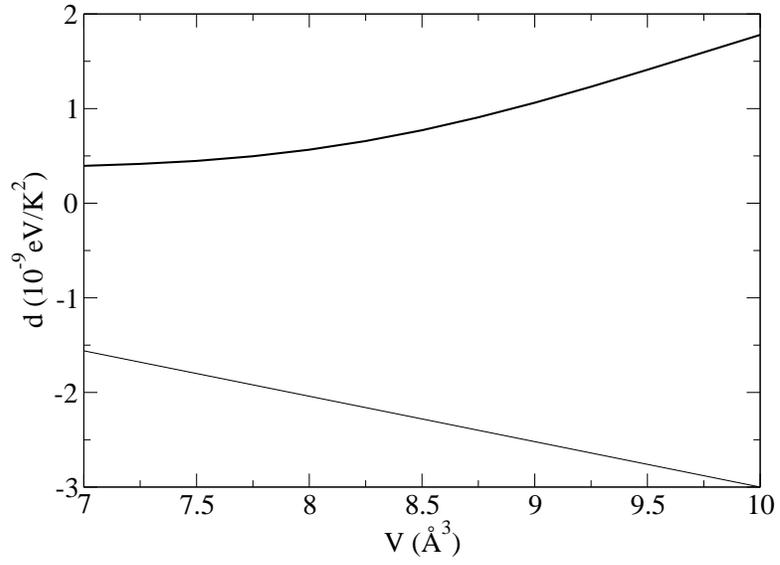}
    \caption{Anharmonic coefficient \(d\) as a function of atomic volume for
      the current PIC calculations (heavy curve) and the vibrationally
      correlated calculations of \citet{apg:2001} (light curve).}
    \label{anharm}
  \end{figure}

  We report in Fig. \ref{anharm} the volume dependence of
  $d$ for $q = 1.60$, compared with the corresponding
  anharmonicity coefficient obtained from the
  vibrationally correlated
  results of \citet{apg:2001}. The coefficients of this
  fit are as follows: \(d^{(0)}=-6.556\times 10^{-8}\),
  \(d^{(1)}=3.399\times 10^{-8}\), \(d^{(2)}=-6.487\times 10^{-9}\),
  \(d^{(3)}=5.400\times 10^{-10}\), \(d^{(4)}=-1.639\times 10^{-11}\),
  where \(d(v)=\sum_{i=0}^4 d^{(i)}V^i\), \(d\) being in eV~K\(^{-2}\)
  and \(V\) in \AA\(^3\). Corresponding data is given in \citet{apg:2001}.
  The striking and important feature of this comparison
  is that the PIC anharmonic free energy is positive,
  whereas the results of \citet{apg:2001} show that in
  reality it is negative. This means that for high-$p$
  hcp Fe, the PIC approximation gives a completely
  incorrect account of anharmonicity, the electronic structure techniques
  in both works being identical. However,
  we note that $f_{\rm anharm}^{\rm PIC}$ is 
  in any case very small. For example, at $V = 7$ \AA$^3$/atom,
  $T = 6000$ K, $f_{\rm anharm}^{\rm PIC} \approx 15$ meV/atom.
  Because it is so small, we ignore its dependence on $q$
  in the following, and use a polynomial fit to the $d(V)$ results
  for $q = 1.60$.
  
  \subsection{Temperature dependence of $c/a$ ratio}
  The equilibrium value $q_{\rm eq}$ of the $c/a$ ratio $q$ at given
  $V$ and $T$ is obtained by minimising the total
  free energy $F_{\rm AI}^{\rm PIC}$ with respect
  to $q$. The variation of $q$ with $T$ comes
  from the $T$- and $q$-dependence of the
  perfect-lattice free energy $U_{\rm perf}^{\rm PIC}$,
  for which we presented results in Sec. \ref{res_perf}, and from
  the $q$-dependence of the mean vibrational
  frequency $\bar{\omega}$ (Eqn. (\ref{f_harm})). (Since
  we neglect the $q$-dependence of the anharmonic
  component of free energy, anharmonicity
  does not contribute to the $T$-dependence of
  $q_{\rm eq}$ here.)
  
\begin{figure}[h]
    \includegraphics{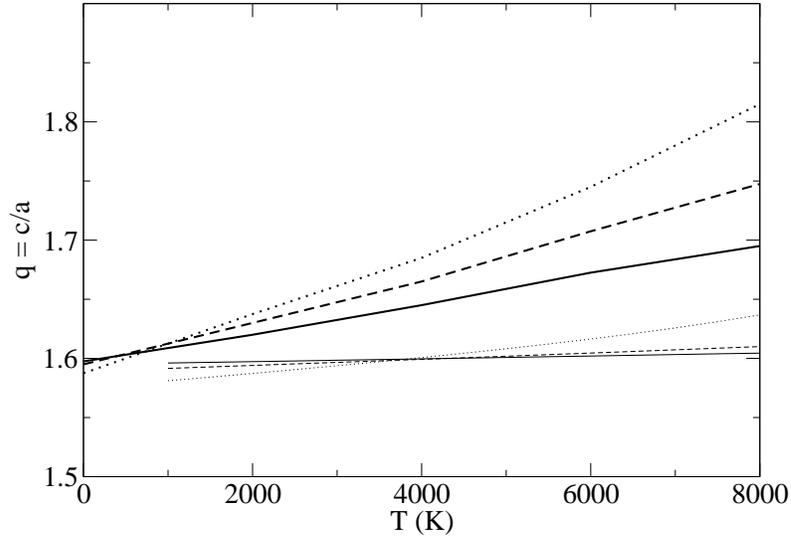}
    \caption{Equilibrium axial ratio \(q\), for \citet{sscg:2001}
      (heavy curves) at atomic volumes of 6.81\AA\(^3\) (solid curve), 7.11\AA\(^3\)
      (dashed curve) and 7.41\AA\(^3\) (dotted curve), and for the current work
      (light curves) at 5.5\AA\(^3\) (solid curve) 7.5\AA\(^3\) (dashed curve) and
      10.0\AA\(^3\) (dotted curve)}
    \label{ratio_tot}
  \end{figure}

  Our results for $q_{\rm eq}$ at three different
  volumes are reported in Fig. \ref{ratio_tot}, where we compare
  them with the earlier PIC predictions of \citet{sscg:2001,ssc:2002}.
  The present results are very
  different from the earlier ones. At all volumes,
  we find only a very weak increase of $q$ with $T$,
  which is between 5 and 10 times smaller than the
  variation reported in \citet{sscg:2001}. We note that
  this gross discrepancy can come only from a
  difference in the $q$-dependence of the harmonic
  mean frequency $\bar{\omega}$. The reason for
  this is that the roughly linear dependence of
  $q_{\rm eq}$ on $T$ seen in both our results
  and those of \citet{sscg:2001} cannot originate
  either from thermal electronic excitations or from
  anharmonicity, since both of these free-energy
  components vary as $T^2$.  The cause of the
  discrepancy, and its implications for understanding
  the elastic anisotropy of the inner core, are
  discussed further in Sec. \ref{discuss}.
  
  \subsection{Thermodynamic functions}
  
  All thermodynamic functions can be calculated by
  taking appropriate derivatives of the PIC free
  energy $F_{\rm AI}^{\rm PIC}$, with
  its perfect-lattice, harmonic and anharmonic
  components parameterised as described above.
  All the results to be presented include the
  dependence of the equilibrium $c/a$ ratio
  $q_{\rm eq}$ on $V$ and $T$. Our
  PIC predictions will be compared with the
  earlier PIC results of \citet{wsc:1996} and
  with the vibrationally correlated results of \citet{apg:2001}.
  
\begin{figure}[h]
    \includegraphics{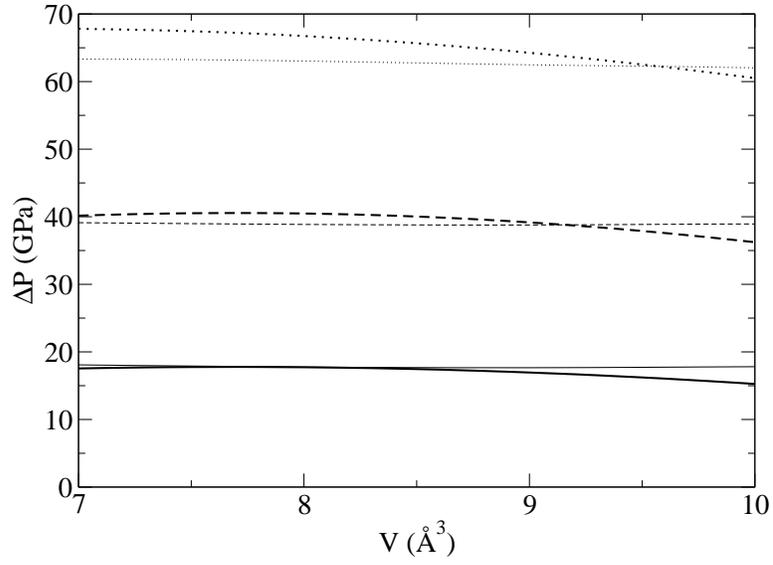}
    \caption{The total thermal pressure on isotherms in this work (light curves)
      and that of \citet{apg:2001} (heavy curves) at 2000~K (solid), 4000~K
      (dashed) and 6000~K (dotted) as a function of atomic volume.}
    \label{tpress}
  \end{figure}

  We begin by considering the thermal pressure
  $\Delta p$, which is one of the most basic
  quantities in interpreting the properties of
  the inner core. This is defined as the difference
  between the pressure in the material at a given
  $V$ and $T$ and the pressure at the same
  volume but at zero temperature:
  \begin{equation}
    \Delta p = - ( \partial F_{\rm AI}^{\rm PIC} /  \partial V )_T +
    ( \partial F_{\rm AI}^{\rm PIC} / \partial V)_{T = 0} \; .
  \end{equation}
  Fig. \ref{tpress} shows a comparison of our PIC results for
  $\Delta p$ with the vibrationally correlated results of \citet{apg:2001}
  on isotherms at $T = 2000$, 4000 and
  6000 K. The rather close agreement shows that the
  PIC approximation gives an accurate account
  of the thermal pressure. This is expected from
  our results for $\bar{\omega}$, since in the harmonic
  approximation $\Delta p$ is given by the electronic
  thermal pressure, which is exactly the same in both
  calculations, plus a vibrational contribution
  equal to $- 3 k_{\rm B} T \partial \ln \bar{\omega} /
  \partial V$. We have seen that for hcp Fe, the PIC
  value of $\ln \bar{\omega}$ differs from the exact
  value by an almost constant offset, which has
  no effect on the derivative of $\ln \bar{\omega}$
  with respect to $V$. Small differences between the
  PIC and vibrationally correlated thermal pressure
  may be due to small variations in the offset between
  the values of \(\ln \bar{\omega}\).
  
  The situation is similar for the thermal energy
  $\Delta E$, defined as the difference between
  the internal energy $E$ at a given $V$ and $T$
  and the internal energy at the same $V$ but
  zero temperature: $\Delta E = E(V,T) - E(V,0)$.
  The electronic part of the thermal energy is identical
  in both PIC and vibrationally correlated calculations,
  and the harmonic vibrational energy is exactly 
  $3 k_{\rm B} T$/atom in both cases, so that any
  difference arises only from the small anharmonic
  contribution. Results for the thermal energy are
  not presented here.
  
\begin{figure}[h]
    \includegraphics{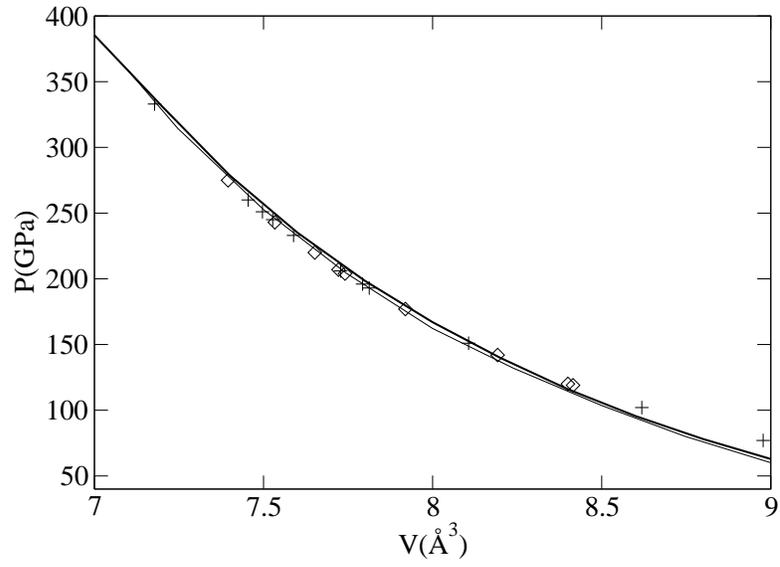}
    \caption{The pressure-volume Hugoniot. Heavy and light curves correspond
      to this work and \citet{apg:2001} respectively; symbols show
      the experimental results of \citet{bmq:1986}.}
    \label{hugp}
  \end{figure}

\begin{figure}[h]
    \includegraphics{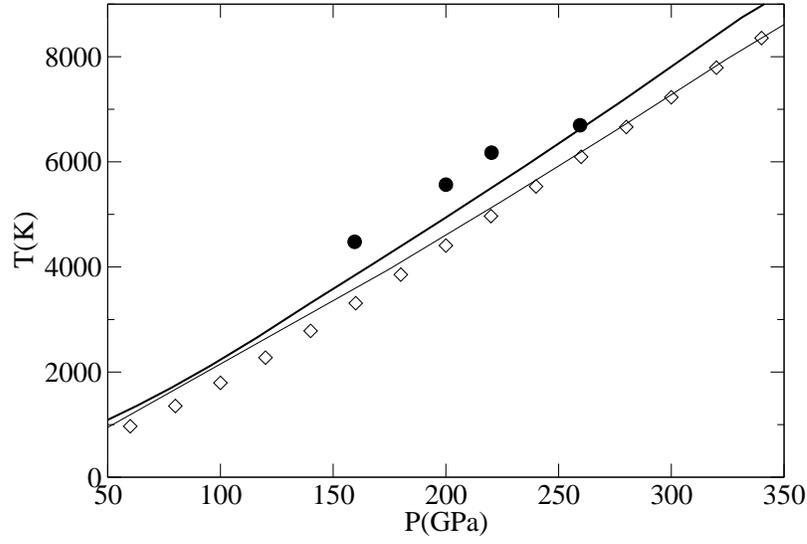}
    \caption{The temperature-pressure Hugoniot. Heavy and light curves correspond
      to this work and \citet{apg:2001} respectively; black circles show the
      experimental results of \citet{yoo:1993} and empty circles are estimates
      due to \citet{bmq:1986}.}
    \label{hugt}
  \end{figure}

  The good accuracy of PIC for thermal pressure
  and thermal energy explains why it has also been
  found to give a good account of shock
  measurements (\citet{wsc:1996}). A conventional
  sequence of shock experiments on samples in the 
  same initial state generates a path through
  thermodynamic state-space known as
  the Hugoniot. On this path, the pressure $p$,
  volume $V$ and internal energy $E$ are related
  by the Rankine-Hugoniot formula (\citet{poirier:2000}):
  \begin{equation}
    \label{rh_eqn}
    \frac{1}{2} ( p + p_0 ) ( V_0 - V ) = E - E_0 \; ,
  \end{equation}
  where $p_0$, $V_0$ and $E_0$ refer to the initial
  state, which usually corresponds to ambient
  conditions. For given $V$, the Hugoniot
  $p$ and temperature $T$ can be determined
  by going to the calculated $p ( V, T )$ and $E ( V, T )$ functions
  and seeking the value of $T$ for which $p$ and $E$ satisfy
  eqn (\ref{rh_eqn}). Fig. \ref{hugp} shows our calculated PIC 
  $p(V)$ on the Hugoniot compared with the experimental results and with the
  fully correlated results
  of \citet{apg:2001}. The almost exact agreement between the
  three sets of results confirms the excellence of PIC for this particular
  purpose. Similar comparisons are shown for the $T(p)$ relation
  on the Hugoniot in Fig. \ref{hugt}.  Here there is a discrepancy,
  which corresponds to a difference in internal energy between the two
  theoretical models.  This difference is fully accounted for by the
  neglect of temperature dependence of \(\bar{\omega}\) in the present work.
  
  We conclude the presentation of results by examining
  three thermodynamic quantities for which our
  comparisons between vibrationally correlated calculations
  and the earlier PIC results of \citet{wsc:1996} and
  \citet{swc:1997} revealed significant differences (\citet{apg:2001}).
  
\begin{figure}[h]
    \includegraphics{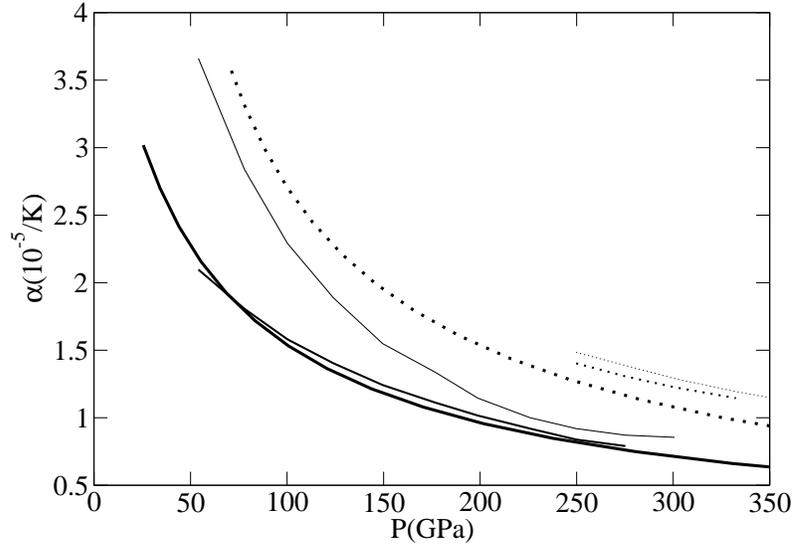}
    \caption{Thermal expansivity on isotherms at 2000~K (solid) and
      6000~K (dotted curves).  Heavy, medium and light curves correspond to this
      work, \citet{apg:2001} and the earlier PIC results of \citet{swc:1997}
      and \citet{wsc:1996} respectively.}
    \label{expans}
  \end{figure}

\begin{figure}[h]
    \includegraphics{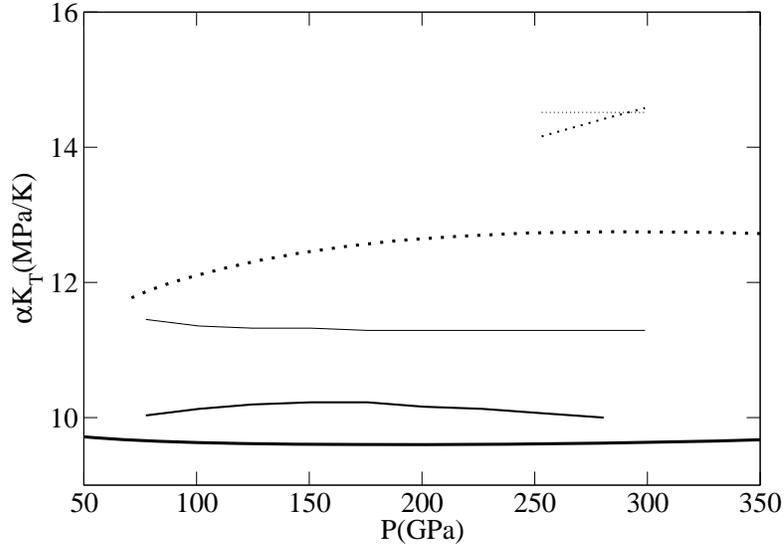}
    \caption{The product \(\alpha K_T\) on isotherms at 2000~K (solid)
      and 6000~K (dotted curves).  Heavy, medium and light curves correspond
      to this work, \citet{apg:2001} and the earlier PIC results of
      \citet{swc:1997} and \citet{wsc:1996} respectively.}
    \label{product}
  \end{figure}

  Fig. \ref{expans} compares the present PIC results
  for thermal expansivity $\alpha$ with earlier calculations.
  At 2000 K, the present results are in very close agreement
  with the vibrationally correlated results, and in strong disagreement
  with the earlier PIC results at low pressures. At higher $T$,
  the present results fall somewhat below both previous sets
  of results.
  The product $\alpha K_T$ (Fig \ref{product}) of expansivity and isothermal
  bulk modulus is important in high-pressure work, because it
  can sometimes be assumed to be independent of $p$ and $T$
  over a wide range of conditions. The vibrationally correlated
  calculations showed that for high-$p$/high-$T$ hcp Fe constancy
  with $p$ is a good approximation, but constancy with $T$ is not.
  The present PIC results show a reduction of $\alpha K_T$ with
  respect to vibrationally correlated and earlier PIC results of,
  at most, around 15\% at 6000K.
  
\begin{figure}[h]
    \includegraphics{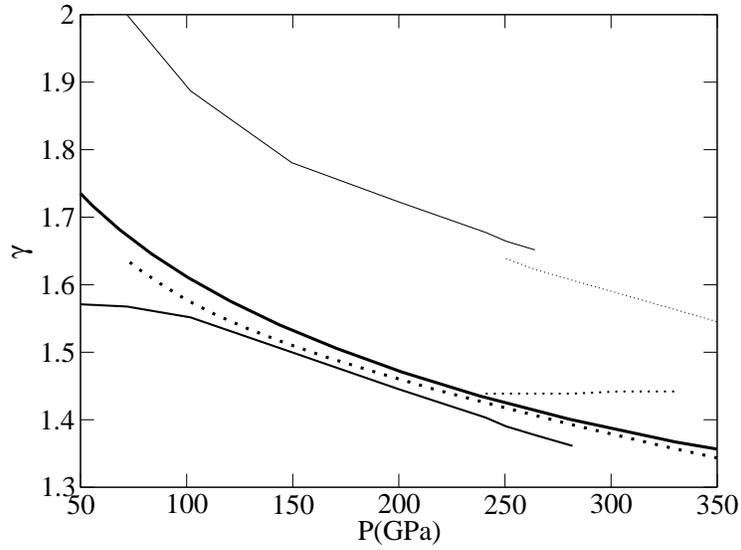}
    \caption{The Gr\"{u}neisen parameter on isotherms at 2000~K (solid)
      and 6000~K (dotted curves).  Heavy, medium and light curves correspond
      to this work, \citet{apg:2001} and the earlier PIC results of \citet{swc:1997}
      and \citet{wsc:1996} respectively.}
    \label{grunie}
  \end{figure}

  The thermodynamic Gr\"{u}neisen parameter $\gamma =
  V ( \partial p / \partial E )_V$  is particularly important,
  because it relates thermal pressure to thermal energy, and
  assumptions about $\gamma$ are often used in reducing
  shock data from Hugoniot to isotherm. The large differences
  between the earlier PIC results and the vibrationally-correlated
  results for $\gamma$ are therefore a cause for concern (\citet{apg:2001}).
  The present PIC results (Fig. \ref{grunie}) agree quite closely with the
  vibrationally-correlated results, and this suggests that the
  cause of the earlier disagreement was not the PIC approximation itself.
  
  \section{Discussion}
  \label{discuss}
  Our comparisons with more exact calculations have shown that the PIC
  approximation gives very good results for a range of important
  thermodynamic properties of hcp Fe at Earth's core conditions.
  A noteworthy example of this is that \(p(V)\) on the Hugoniot
  agrees almost perfectly with the more exact results. Our analysis
  of the free energy into different components makes clear why 
  PIC is so good. The perfect-lattice component is exactly the
  same in the two approaches. For the harmonic component,
  the sole requirement for good results is that the logarithmic
  derivative $d \ln \bar{\omega} / d \ln V$ of the geometric
  mean frequency $\bar{\omega}$ be correct. But we have
  seen that for hcp Fe the PIC $\bar{\omega}$ differs
  from the $\bar{\omega}$ given by calculation of the
  full phonon spectrum by an almost constant factor over a wide
  range of volumes, so that this requirement is satisfied. The basic
  reason for this is that the phonon dispersion relations of hcp Fe scale by
  a uniform factor with changing volume (see Fig. 3 of
  \citet{apg:2001}).
  Finally, the anharmonic component of free energy is very small,
  and has only a very minor effect on most thermodynamic
  functions. The reliability of PIC actually requires that 
  anharmonic effects be small, since we have shown that the
  PIC approximation misrepresents these effects
  in predicting the wrong sign of the anharmonic
  free energy.
  
  Surprisingly, even though PIC appears to be so good, we
  find important discrepancies with the earlier PIC results
  of \citet{swc:1997} and \citet{wsc:1996}. In particular, our calculations
  of the
  thermodynamic Gr\"{u}neisen parameter agree much more closely with the
  calculations of \citet{apg:2001}.  These discrepancies are clearly not
  due to PIC itself, but must come from other technical differences.
  We note that in the work of \citet{wsc:1996}, the PIC calculations
  actually employed a tight-binding representation of the
  total energy function, the parameters in the tight-binding
  model being fitted to {\em ab initio\/} calculations. Conceivably,
  the tight-binding fit might have led to errors.
  
  Even more surprising is that the strong increase with temperature
  of the axial $c/a$ ratio predicted by recent PIC calculations
  is not reproduced at all by the present PIC work. According
  to \citet{sscg:2001}, at the atomic volume of 7.11 \AA$^3$, $c/a$
  increases from 1.63 to 1.75 as $T$ goes from 2000 to 8000 K, whereas
  in the present PIC calculations at the similar volume of 7.0 \AA$^3$,
  $c/a$ increases only from 1.594 to 1.610 over the
  same temperature range. The correctness
  of the weak $c/a$ variation found here is supported by preliminary
  calculations (\citet{alfe_ca}) using the techniques of
  \citet{apg:2001,apg:2002}.  The reasons
  for this discrepancy are completely unclear at present. The discrepancy
  has major implications for our understanding of the Earth's
  inner core, because the recently proposed re-interpretation of the
  elastic anisotropy of the inner core appears to depend crucially
  on a strong $T$ variation of $c/a$. We believe it is highly desirable
  that this question be re-examined by other research groups.
  
  Our work sheds light on the usefulness of the PIC approximation. Since
  we have seen that PIC cannot be relied on for anharmonic
  contributions, it should be regarded as a way of calculating the
  geometric-mean harmonic frequency $\bar{\omega}$. But PIC
  requires {\em ab initio\/} calculations on a periodic system in which
  a single atom is displaced from its perfect-lattice site. These are
  precisely the same calculations that are performed in order to
  obtain the force-constant matrix used to compute the full phonon
  spectrum (\citet{apg:2001,apg:2002}). For an {\em ab initio\/} method that
  yields forces
  on all ions -- which includes the pseudopotential and
  PAW implementations of DFT, among others -- the net result of
  the PIC procedure is to discard all the information contained
  in the ionic forces, retaining only the variation of total energy
  with displacement. This suggests that it may be better to use the force
  information to compute the force-constant matrix and hence the full phonon
  spectrum, rather than adopting the PIC approach.
  
  \bibliography{all_papers}

\begin{thebibliography}{43}
\expandafter\ifx\csname natexlab\endcsname\relax\def\natexlab#1{#1}\fi
\expandafter\ifx\csname url\endcsname\relax
  \def\url#1{\texttt{#1}}\fi
\expandafter\ifx\csname urlprefix\endcsname\relax\def\urlprefix{URL }\fi

\bibitem[{Alf\'e(2002)}]{alfe_ca}
Alf\'e, D., 2002. Unpublished.

\bibitem[{Alf{\`e} et~al.(2000)Alf{\`e}, Kresse, and Gillan}]{akg:2000}
Alf{\`e}, D., Kresse, G., Gillan, M.~J., 2000. Structure and dynamics of liquid
  iron under \uppercase{E}arth's core conditions. Phys. Rev. B 61, 132.

\bibitem[{Alf\`{e} et~al.(2001)Alf\`{e}, Price, and Gillan}]{apg:2001}
Alf\`{e}, D., Price, G.~D., Gillan, M.~J., 2001. Thermodynamics of
  hexagonal-close-packed iron under \uppercase{E}arth's core conditions. Phys.
  Rev. B 64, 045123.

\bibitem[{Alf\`{e} et~al.(2002)Alf\`{e}, Price, and Gillan}]{apg:2002}
Alf\`{e}, D., Price, G.~D., Gillan, M.~J., 2002. Iron under \uppercase{E}arth's
  core conditions: Liquid-state thermodynamics and high-pressure melting curve
  from ab initio calculations. Phys. Rev. B 65, 165118.

\bibitem[{Belonoshko et~al.(2000)Belonoshko, Ahuja, and Johansson}]{baj:2000}
Belonoshko, A.~B., Ahuja, R., Johansson, B., 2000. Quasi-\textit{ab initio}
  molecular dynamic study of \uppercase{F}e melting. Phys. Rev. Lett. 84, 3638.

\bibitem[{Bl{\"o}chl(1994)}]{blochl:1994}
Bl{\"o}chl, P.~E., 1994. Projector augmented-wave method. Phys. Rev. B 50, 953.

\bibitem[{Brodholt et~al.(2002)Brodholt, Oganov, and Price}]{bop:2002}
Brodholt, J., Oganov, A., Price, D., 2002. Computational mineral physics and
  the physical properties of perovskite. Phil. Trans. Roy. Soc. 360, 2507.

\bibitem[{Brown and McQueen(1986)}]{bmq:1986}
Brown, J.~M., McQueen, R.~G., 1986. Phase-transitions,
  \uppercase{G}r{\"u}neisen-parameter and elasticity for shocked iron between
  77-\uppercase{GP}a and 400-\uppercase{GP}a. J. Geophys. Res., [Space Phys]
  91, 7485.

\bibitem[{Cohen and G{\"u}lseren(2002)}]{gc_tant:2002}
Cohen, R.~E., G{\"u}lseren, O., 2002. High-pressure thermoelasticity of
  body-centered-cubic tantalum. Phys. Rev. B 65, 064103.

\bibitem[{Cowley et~al.(1990)Cowley, Gross, Gong, and Horton}]{c_ea:1990}
Cowley, E.~R., Gross, J., Gong, Z.~X., Horton, G.~K., 1990. Cell-cluster and
  self-consistent calculations for a model sodium-chloride crystal. Phys. Rev.
  B 42, 3135.

\bibitem[{Creager(1992)}]{creager:1992}
Creager, K.~C., 1992. Anisotropy of the inner core from differential
  travel-times of the phases pkp and pkikp. Nature 356, 309.

\bibitem[{Frenkel and Smit(1996)}]{frenk:1996}
Frenkel, D., Smit, B., 1996. Understanding Molecular Simulation. Academic, San
  Diego.

\bibitem[{Gillan(1989)}]{gillan:1989}
Gillan, M.~J., 1989. Calculation of the vacancy formation energy in aluminium.
  J. Phys.: Condens. Matter 1, 689.

\bibitem[{G{\"u}lseren and Cohen(2001)}]{cg_tant:2001}
G{\"u}lseren, O., Cohen, R.~E., 2001. Thermal equation of state of tantalum.
  Phys. Rev. B 63, 224101.

\bibitem[{Hirshfelder et~al.(1954)Hirshfelder, Curtiss, and Bird}]{hcb:1954}
Hirshfelder, J.~O., Curtiss, C.~F., Bird, R.~B., 1954. Molecular Theory of
  Gases and Liquids. John Wiley and Sons, Inc, New York.

\bibitem[{Hohenberg and Kohn(1964)}]{hk:1964}
Hohenberg, P., Kohn, W., 1964. Inhomogeneous electron gas. Phys. Rev. 136,
  B864.

\bibitem[{Holt et~al.(1970)Holt, Hoover, Gray, and Shortle}]{h_ea:1970}
Holt, A.~C., Hoover, W.~G., Gray, S.~G., Shortle, D.~R., 1970. Comparison of
  the lattice-dynamics and cell-model approximations with
  \uppercase{M}onte-\uppercase{C}arlo thermodynamic properties. Physica 49, 61.

\bibitem[{Holt and Ross(1970)}]{hr:1970}
Holt, A.~C., Ross, M., 1970. Calculations of the \uppercase{G}r{\"u}neisen
  parameter of some models of the solid. Phys. Rev. B 1, 2700.

\bibitem[{Jones and Gunnarsson(1989)}]{jg:1989}
Jones, R.~O., Gunnarsson, O., 1989. The density functional formalism, its
  applications and prospects. Rev. Mod. Phys. 61, 689.

\bibitem[{Kohn and Sham(1965)}]{ks:1965}
Kohn, W., Sham, L., 1965. Self-consistent equations including exchange and
  correlation effects. Phys. Rev. 140, A1133.

\bibitem[{Kresse and Furthm{\"u}ller(1996{\natexlab{a}})}]{kf_02:1996}
Kresse, G., Furthm{\"u}ller, J., 1996{\natexlab{a}}. Efficiency of \textit{ab
  initio} total energy calculations for metals and semiconductors using a
  plane-wave basis set. Comput. Mater. Sci. 6, 15.

\bibitem[{Kresse and Furthm{\"u}ller(1996{\natexlab{b}})}]{kf_01:1996}
Kresse, G., Furthm{\"u}ller, J., 1996{\natexlab{b}}. Efficient iterative
  schemes for \textit{ab initio} total-energy calculations using a plane-wave
  basis set. Phys. Rev. B 54, 11169.

\bibitem[{Kresse and Joubert(1999)}]{kj:1999}
Kresse, G., Joubert, D., 1999. From ultrasoft pseudopotentials to the projector
  augmented-wave method. Phys. Rev. B 59, 1758.

\bibitem[{Laio et~al.(2000)Laio, Bernard, Chiarotti, Scandolo, and
  Tosatti}]{lbcst:2000}
Laio, A., Bernard, S., Chiarotti, G.~L., Scandolo, S., Tosatti, E., 2000.
  Physics of iron at \uppercase{E}arth's core conditions. Science 287, 1027.

\bibitem[{Mermin(1965)}]{mermin:1965}
Mermin, N.~D., 1965. Thermal properties of the inhomogeneous electron gas.
  Phys. Rev. 137, A1441.

\bibitem[{Monkhorst and Pack(1976)}]{mp:1976}
Monkhorst, H.~J., Pack, J.~D., 1976. Special points for
  \uppercase{B}rillouin-zone integrations. Phys. Rev. B 13, 5188.

\bibitem[{Oganov et~al.(2001)Oganov, Brodholt, and Price}]{obp:2001}
Oganov, A.~R., Brodholt, J.~P., Price, G.~D., 2001. The elastic constants of
  \uppercase{M}g\uppercase{S}i\uppercase{O}\(_3\) perovskite at pressures and
  temperatures of the \uppercase{E}arth's mantle. Nature 411, 934.

\bibitem[{Perdew et~al.(1992)Perdew, Chevary, Vosko, Jackson, Pederson, Singh,
  and Fiolhais}]{perdew:1992}
Perdew, J.~P., Chevary, J.~A., Vosko, S.~H., Jackson, K.~A., Pederson, M.~R.,
  Singh, D.~J., Fiolhais, C., 1992. Atoms, molecules, solids and surfaces -
  applications of the generalized gradient approximation for exchange and
  correlation. Phys. Rev. B 46, 6671.

\bibitem[{Poirier(2000)}]{poirier:2000}
Poirier, J.-P., 2000. Introduction to the Physics of the \uppercase{E}arth's
  Interior, 2nd Edition. Cambridge University Press, Cambridge.

\bibitem[{Ree and Holt(1973)}]{rh:1973}
Ree, F.~H., Holt, A.~C., 1973. Thermodynamic properties of the alkali-halide
  crystals. Phys. Rev. B 8, 826.

\bibitem[{Song and Helmberger(1993)}]{sh:1993}
Song, X.~D., Helmberger, D.~V., 1993. Anisotropy of \uppercase{E}arth's
  inner-core. Geophys. Res. Lett. 20, 2591.

\bibitem[{Steinle-Neumann et~al.(1999)Steinle-Neumann, Stixrude, and
  Cohen}]{ssc:1999}
Steinle-Neumann, G., Stixrude, L., Cohen, R.~E., 1999. First-principles elastic
  constants for the hcp transition metals \uppercase{F}e,\uppercase{C}o, and
  \uppercase{R}e at high pressure. Phys. Rev. B 60, 791.

\bibitem[{Steinle-Neumann et~al.(2002)Steinle-Neumann, Stixrude, and
  Cohen}]{ssc:2002}
Steinle-Neumann, G., Stixrude, L., Cohen, R.~E., 2002. Physical properties of
  iron in the inner core. In: Dehant, V., Creager, K., Zatman, S., Karato,
  S.-I. (Eds.), Core structure, dynamics and rotation. American Geophysical
  Union, Washington, DC, pp. 137--161.

\bibitem[{Steinle-Neumann et~al.(2001)Steinle-Neumann, Stixrude, Cohen, and
  G{\"u}lseren}]{sscg:2001}
Steinle-Neumann, G., Stixrude, L., Cohen, R.~E., G{\"u}lseren, O., 2001.
  Elasticity of iron at the temperature of the \uppercase{E}arth's inner core.
  Nature 413~(6851).

\bibitem[{Stixrude and Cohen(1995)}]{sc:1995}
Stixrude, L., Cohen, R.~E., 1995. High-pressure elasticity of iron and
  anisotropy of \uppercase{E}arth's inner-core. Science 267, 1972.

\bibitem[{Stixrude et~al.(1998)Stixrude, Cohen, and Hemley}]{sch:1998}
Stixrude, L., Cohen, R.~E., Hemley, R.~J., 1998. Theory of minerals at high
  pressure. Rev. Mineral. 37, 639.

\bibitem[{Stixrude et~al.(1997)Stixrude, Wasserman, and Cohen}]{swc:1997}
Stixrude, L., Wasserman, E., Cohen, R.~E., 1997. Composition and temperature of
  \uppercase{E}arth's inner core. J. Geophys. Res., [Space Phys] 102, 24729.

\bibitem[{Tromp(1993)}]{tromp:1993}
Tromp, J., 1993. Support for anisotropy of the \uppercase{E}arth's inner-core
  from free oscillations. Nature 366, 678.

\bibitem[{Wang and Perdew(1991)}]{wp:1991}
Wang, Y., Perdew, J.~P., 1991. Correlation hole of the spin-polarized electron
  gas, with exact small-wave-vector and high-density scaling. Phys. Rev. B 44,
  13298.

\bibitem[{Wasserman et~al.(1996)Wasserman, Stixrude, and Cohen}]{wsc:1996}
Wasserman, E., Stixrude, L., Cohen, R.~E., 1996. Thermal properties of iron at
  high pressures and temperatures. Phys. Rev. B 53, 8296.

\bibitem[{Wentzcovitch et~al.(1992)Wentzcovitch, Martins, and Allen}]{wma:1992}
Wentzcovitch, R.~M., Martins, J.~L., Allen, P.~B., 1992. Energy versus
  free-energy conservation in 1st-principles molecular-dynamics. Phys. Rev. B
  45, 11372.

\bibitem[{Westra and Cowley(1975)}]{wc:1975}
Westra, K., Cowley, E.~R., 1975. Cell-cluster expansion for an anharmonic
  solid. Phys. Rev. B 11, 4008.

\bibitem[{Yoo et~al.(1993)Yoo, Holmes, Ross, Webb, and Pike}]{yoo:1993}
Yoo, C.~S., Holmes, N.~C., Ross, M., Webb, D.~J., Pike, C., 1993. Shock
  temperatures and melting of iron at \uppercase{E}arth core conditions. Phys.
  Rev. Lett. 70, 3931.

\end{thebibliography}
  
  \pagebreak
  
  \pagebreak
  
  \pagebreak
  
  \pagebreak
  
  \pagebreak
  
  \pagebreak
  
  \pagebreak
  
  \pagebreak
  
  \pagebreak
  
\end{document}